\documentclass[%
 reprint,
superscriptaddress,
showpacs,preprintnumbers,
 amsmath,amssymb,
 aps,
 prb,
floatfix,
]{revtex4-1}

\usepackage{sidecap}
\sidecaptionvpos{figure}{c}

\usepackage{graphicx}% Include figure files
\usepackage{dcolumn}% Align table columns on decimal point
\usepackage{bm}% bold math
\usepackage{amsmath,amsfonts,
	amsgen,amsbsy,amsopn,amstext,amssymb
}
\usepackage{epstopdf}
\usepackage{color}
\begin{document}

\preprint{}

\title{Bright and Gap Solitons in Membrane-Type Acoustic Metamaterials}

\author{Jiangyi  Zhang} \email{Jiangyi.Zhang.Etu@univ-lemans.fr}
\author{Vicente Romero-Garc{\'i}a}   
\author{Georgios Theocharis}  
\author{Olivier  Richoux } 
\author{Vassos Achilleos} 
\affiliation{Laboratoire d'Acoustique de l'Universit{\'e} du Maine - CNRS UMR 6613, Le Mans, France}

\author{Dimitrios J. Frantzeskakis}
\affiliation{Department of Physics, National and Kapodistrian University of Athens, 
	Panepistimiopolis, Zografos, Athens 15784, Greece}

\date{\today}

\begin{abstract}

We study analytically and numerically envelope solitons (bright and gap solitons) in a one-dimensional, nonlinear acoustic metamaterial, composed of an air-filled waveguide periodically loaded by clamped elastic plates. Based on the transmission line approach, we derive a nonlinear dynamical lattice model which, in the continuum approximation, leads to a nonlinear, dispersive and dissipative wave equation. 
Applying the multiple scales perturbation method, we derive
an effective lossy nonlinear Schr\"odinger equation and obtain analytical 
expressions for bright and gap  solitons.
We also perform direct numerical simulations to 
%numerically 
study the dissipation-induced dynamics of the bright and gap solitons. 
%and we compare it  with analytical results obtained by soliton perturbation theory. 
Numerical and analytical results, relying on the analytical approximations and perturbation theory 
for solions, are  found to be in good agreement. 
%confirming that the system supports envelope solitary waves which are robust in the presence of propagating viscothermal losses.

%\begin{description}
%\item[PACS numbers]May be entered using the \verb+\pacs{#1}+ command.
%You may use the \texttt{description} environment to structure your abstract;use the optional argument of the \verb+\item+ command to give the category of each item. 
%\end{description}
\end{abstract}

\maketitle

%\tableofcontents

\section{Introduction}

Acoustic metamaterials, namely structured materials made of resonant building blocks, 
present strong dispersion around the resonance frequency. 
In acoustic waveguides, this resonance-induced dispersion was observed 
for the first time by Sugimoto\cite{ref-Sugimoto} and Bradley\cite{ref-Bradley}. 
Later, Liu {\it et al.}\cite{ref-Liu} paved the way for the realization of 
acoustic metamaterials, through arrangements of locally resonant elements, 
that could be described as effective media with negative effective parameters, 
%properties 
not found in natural materials. Since then, a plethora of exotic properties 
of acoustic metamaterials have been intensively exploited showing 
novel wave control phenomena; these include 
%such as 
subwavelength focusing\cite{Koshevich10}, cloaking \cite{Dehesa13}, perfect absorption \cite{Ma15,Romero16} and extraordinary transmission\cite{Park13} among others\cite{ref-Deymier}.

%Key features playing crucial role in the propagation of waves through a medium are 
Generally, dispersion, nonlinearity and dissipation play a key role in wave propagation, 
with all these phenomena appearing generically in practice.
%which in practice are unavoidable. 
However, in acoustic metamaterials --and up to now-- only few works have systematically consider 
%tried to study 
the interplay between all the above 
%effects 
phenomena\cite{ref-Naugolnykh,ref-Bradley-Nonlinear,ref-Sugimoto1,ref-SugimotoJASA,ref-Richoux}. 
Particularly, in some works, the combined effects of dissipation and dispersion 
%properties was 
were studied without considering the nonlinearity\cite{ref-zwikker,ref-solymar,ref-Theocharis,ref-Bradley,ref-Bradley2}; in this case, the relevance of dissipation was further exploited 
%the dissipative effect was exploited further 
to the design of perfect absorbers\cite{Ma15,Romero16}. 
In fact, the majority of 
%most of the  
works on acoustic metamaterials focus on the linear regime and do not 
consider the nonlinear response of the structure.
Nevertheless, due to the intrinsically nonlinear nature of the problem and 
%Due to 
the strong dispersion introduced by the locally resonant building blocks, 
%the 
acoustic metamaterials are good candidates to study the combined effects of nonlinearity and dispersion that can give rise to interesting nonlinear 
%several nonlinear wave propagation 
effects. These include the beating of the higher generated harmonics\cite{ref-Morcillo,ref-Noe,ref-us}, 
self demodulation\cite{ref-Hamilton,ref-Michalakis}, as well as the emergence 
%existence 
of solitons\cite{ref-Remoissenet,ref-Sugimoto1,ref-Achilleos}, namely robust localized waves propagating undistorted due to a balance between dispersion and nonlinearity.
%which may give rise to different phenomena such as harmonic generation or self demodulation among others\cite{ref-Hamilton}. 

In this work, we show the existence and investigate the dynamics of bright and gap solitons 
in an acoustic metamaterial composed of an air-filled waveguide, %considering viscothermal losses, 
periodically loaded by clamped elastic plates, taking into regard viscothermal losses. 
Based on the transmission line (TL) approach used widely in acoustics  
\cite{LissekPRB,acousticTL_1,acousticTL_2,acousticTL_3,ref-Bongard}, we derive a nonlinear 
%dissipative  
dynamical lattice model which, in the continuum approximation, leads to 
a nonlinear dispersive and dissipative wave equation. By applying the multiple scales 
perturbation method, we find that the evolution of the pressure can be described by a 
%{\it lossy} 
nonlinear Schr\"odinger (NLS) equation incorporating linear loss. %As a consequence, 
We thus show that --in the lossless case-- the system supports envelope solitons, 
which are found in a closed analytical form. We perform direct numerical simulations, 
in the framework of the nonlinear 
%dynamical 
lattice model, and show that both bright and gap solitons are supported in the system; 
numerical results are found to be in good agreement with the analytical solutions of 
the NLS equation. In addition, we numerically and analytically study the effect of 
viscothermal losses on the envelope solitons, showing that, for the particular 
setting considered herein, the soliton solutions are less affected by the losses 
than dispersion.

The paper is structured as follows. In Section \ref{sec2}, we introduce our setup and derive 
the one-dimensional (1D) nonlinear dissipative lattice model, %and derive 
as well as the nonlinear dissipative and dispersive wave equation and the 
associated dispersion relation. In Section \ref{sec3}, 
%by applying a multiple scale perturbation method, 
we derive the lossy NLS equation via the multiple scale perturbation method. 
%present its soliton solutions in the absence of losses. 
Then, we start by studying the dynamics of 
%envelope 
bright solitons, both in the lossless and lossy cases, and we 
%end 
complete our investigations by presenting the dynamics of gap solitons.
Finally, Section \ref{sec5} summarizes our findings and discusses future research directions.

\section{Electro-Acoustic Analogue Modeling}\label{sec2}

\subsection{Setup and model}

\begin{figure}[tbp]
\centering
\includegraphics[width=8.5cm]{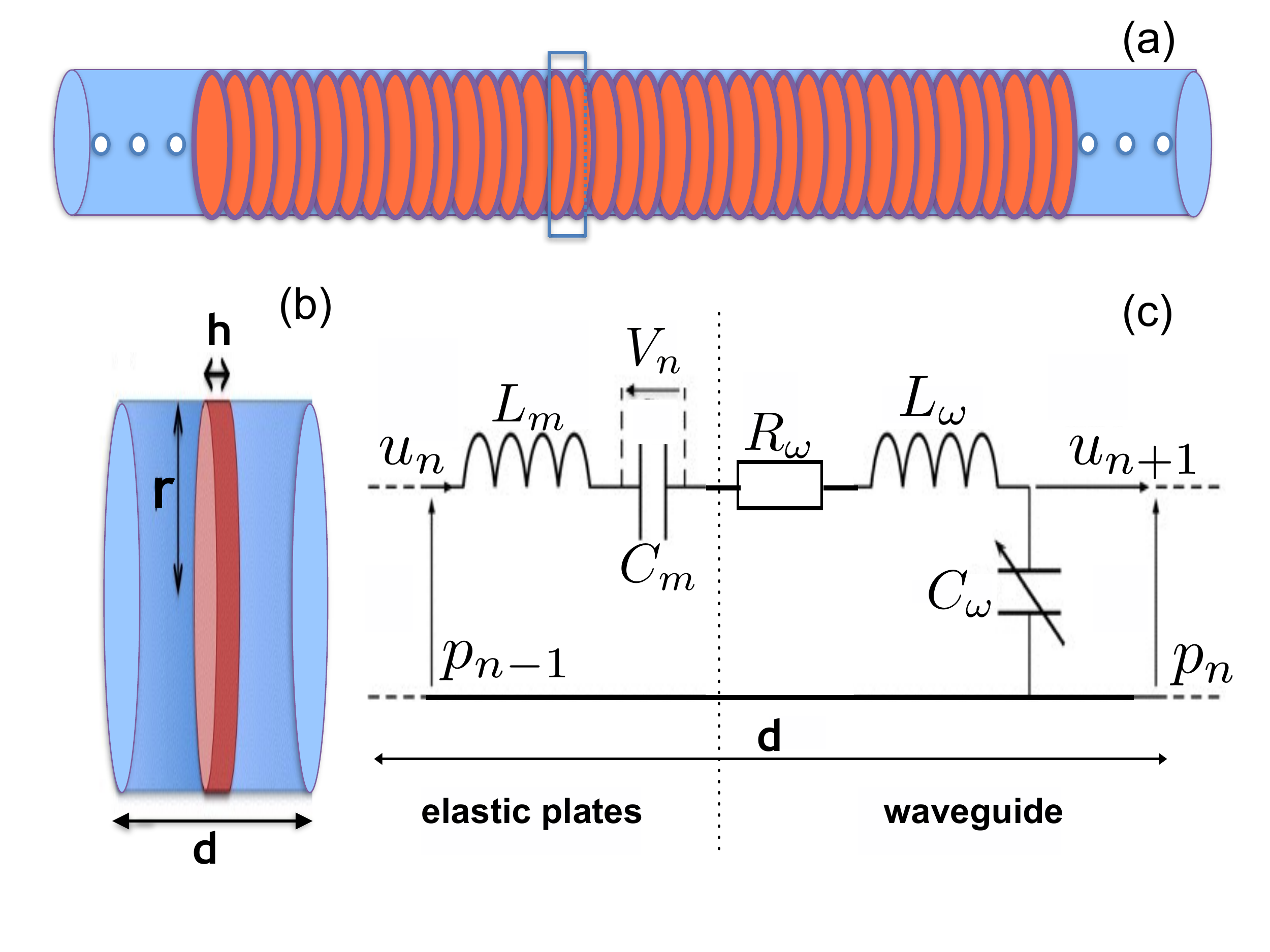} 
\caption{(Color online) (a) Waveguide loaded with an array of elastic plates; (b) the unit-cell 
of the system; (c) corresponding unit-cell circuit.}
\label{fig:fig1}
\end{figure}

A schematic view of the acoustic waveguide periodically loaded by clamped elastic plates, 
as well as the respective unit-cell structure of this setup 
are respectively shown in Figs.~\ref{fig:fig1}(a) and \ref{fig:fig1}(b). 
We consider low-frequency wave propagation in this setting, 
%structure, 
i.e., the frequency range is well below the first cut-off frequency of the higher propagating 
modes in the waveguide, therefore the problem is considered as one-dimensional (1D). 

In order to theoretically analyze this system, we employ the electro-acoustic analogy; 
this allows us to derive a nonlinear discrete wave equation for an equivalent 
electrical TL, which, in the continuum limit, can be studied 
%analytically 
by means of the method of multiple scales. 
%solved perturbatively in the continuum limit. 
%Such an approach provides an efficient way to treat analytically the problem which, 
Our approach is much simpler than the one relying on the study  
%otherwise, should be studied in the framework 
of a nonlinear acoustic wave equation 
coupled with a set of differential equations describing the dynamics of each 
elastic plate. Furthermore, our approach 
%is much simpler, 
allows for a straightforward analytical treatment of the problem by means of standard techniques 
that are used in other physical systems\cite{ref-Remoissenet}. 

The unit-cell circuit of the equivalent TL model of this setting is shown in 
Fig.~\ref{fig:fig1}(c). It consists of two parts, one corresponding to the propagation 
in the acoustic waveguide, and the other to the elastic plate 
(separated in Fig.~\ref{fig:fig1}(c) by a thin vertical dotted line).
The voltage $v$ and the current $i$ of the equivalent electrical TL 
corresponds to the acoustic pressure $p$ and to the volume velocity $u$ 
flowing through the waveguide cross-section, respectively\cite{ref-Achilleos,LissekPRB}. 
The above 
%modeling is 
considerations are valid in the low frequency regime, i.e., when the wavelength $\lambda \gg d$.
%The unit-cell consists of two parts, one corresponding to the propagation in the acoustic waveguide \sout{(\sout{tube} \textcolor{red}{noted by the subscript w in %Fig. \ref{fig:fig1}(c)})} and the other to the elastic plate \sout{(\textcolor{red}{noted by the subscript m and}} separated in Fig. \ref{fig:fig1}(c) by a thin vertical dotted %line). 

The resonant elastic plate can be modeled by a $LC$ circuit, namely the series 
combination of an inductance $L_{m}=\rho_{m}h/S$ and a capacitance 
$C_{m}=(\omega_{m}^{2} L_{m})^{-1}$, where $\rho_{m}$ is the plate density, 
$S$ represents the cross-section area of the plate, 
while $\omega_{m}=2 \pi f_m$ is the resonance frequency of the plate, with
\begin{equation}
	f_m=0.4694 \frac{h}{r^2} \sqrt{\frac{E}{\rho_m (1-\nu^2)}},
	\label{eq:eq1new}
\end{equation}
\noindent
where $E$ is the Young's modulus and $\nu$ is the Poisson ratio\cite{LissekPRB,ref-Bongard}. Losses originating from the dynamic response of the elastic plates are not taken into account in this work.

The part of the unit-cell circuit that corresponds to the waveguide solely 
(i.e., without the elastic plates and the associated periodic structure) 
is modeled by the inductance $L_{\omega}$, the resistance $R_{\omega}$ 
and shunt capacitance $C_{\omega}$; the linear part of the inductance is 
$L_{\omega 0}=\rho_{0}d/S$ and the capacitance is $C_{\omega 0}=Sd/(\rho_{0} c_{0}^{2})$, 
where $\rho_0$ and $c_0$ are the density and the sound velocity of the fluid 
in the waveguide respectively; the latter, has a cross section $S=\pi r^2$. 
The resistance $R_{\omega}={\rm Im}(kZ_c)d$ (${\rm Im}$ 
%corresponds to 
stands for the imaginary part) corresponds to propagation losses 
due to viscous and thermal effects; here, the wavenumber $k$ is connected 
with the frequency $\omega$ through the following equation \cite{ref-zwikker},
\begin{equation}
k=\frac{\omega}{c_0}\left( 1+\frac{1-i}{s}(1+(\gamma-1)/\sqrt{Pr}) \right),
\end{equation}
while $Z_c$ is given by:
\begin{equation}
 Z_c=\frac{\rho_0 c_0}{S}\left( 1+\frac{1-j}{s}(1-(\gamma-1)/\sqrt{Pr}) \right),
\label{zc}
\end{equation}
\noindent
%
%Therefore, the resistance $R_{\omega}={\rm Im}(kZ_c)d$ (${\rm Im}$ corresponds to the imaginary part), represents a viscothermal loss due to wave propagation in the unit-cell of the system; 
%Here, $Z_c=\sqrt{K\rho}/S$ is the complex characteristic impedance, and $k=\omega\sqrt{\rho/K}$ is the complex wave number, with bulk modulus
%\textcolor{red}{I changed j by i in the following}
%\begin{equation}
%K=\gamma p_0 \left[1+\left(\gamma-1\right)\frac{2}{Bs\sqrt{-i}}\frac{J_1\left(Bs\sqrt{-i}\right)}{J_0\left(Bs\sqrt{-i}\right)}\right]^{-1},
%\label{eq:eq2new}
%\end{equation}
%\noindent
%and density, 
%\begin{equation}
%\rho=\rho_0 \left[1-\frac{2}{s\sqrt{-i}}\frac{J_1\left(s\sqrt{-i}\right)}{J_0\left(s\sqrt{-i}\right)}\right]^{-1}, 
%\label{eq:eq3new}
%\end{equation}
\noindent
where $\gamma$ is the specific heat ratio, $Pr$ is the Prandtl number, 
and $s=\sqrt{\omega\rho_0 r^2/\eta}$, with $\eta$ being the shear viscosity. %is $B^2=0.71$
%
%$s=\sqrt{\omega\rho_0 R^2/\eta}$ \textcolor{red}{R is the radius ?? $r$}, $J_0$ and $J_1$ are Bessel functions %\cite{Stinson91, ref-JFAllard} \textcolor{red}{be carefull with ref}. For wave propagation in air at $18^{\circ}$ C, the shear viscosity is $\eta=1.84\;10^{-5}$ kg/m/s, %the thermal conductivity is \textcolor{red}{change k it's wavenumber after ...} $k=2.6\;10^{-2} $ w/m/k, the specific heat ratio is $\gamma=1.4$, and Prandtl number %is $B^2=0.71$. \\
%
Here, we approximate the frequency dependent viscothermal losses by a resistance with a constant value 
around the frequency of the narrow spectral width envelope solutions that we are interested in. 
%{\color{blue} To summarize, the above modeling is valid in the low frequency regime (sufficiently lower to the Bragg frequency), considering lossless response of the elastic plates and approximating the frequency dependent viscothermal losses by a resistance with a constant value around the frequency of our interest. 

%A final assumption is that we consider 
At this point, we mention that we consider the response 
of the elastic plate to be linear, while the propagation in the waveguide weakly nonlinear. 
This is a reasonable approximation, since the pressure amplitudes used 
in this work are not sufficiently strong to excite nonlinear vibrations of the elastic 
plate\cite{ref-Chandrasekharappa}. On the other hand, it is well known that  
%We know that 
due to the compressibility of air the wave celerity is nonlinear, $c_{NL}$. 
This, in turn, lead us to consider that the capacitance $C_{\omega}$ is nonlinear, 
depending on the pressure $p$, while 
%; on the other hand, 
the inductance is assumed to be linear: $L_{\omega 0}=L_{\omega}$. 
Approximating the celerity as $c_{NL} \approx c_{0}\left(1+\beta_{0}p/\rho_{0} c_{0}^{2} \right)$, 
where $ \beta_{0}$ is the nonlinear parameter ($\beta_{0}=1.2$ for air), 
the pressure-dependent capacitance $C_{\omega}$ can be expressed as 
\begin{equation}
	C_{\omega} \simeq  C_{\omega 0}-C_{\omega}^{'}p_{n}
	\label{eq:eq4new},
\end{equation}
where $C_{\omega 0}=Sd/\varrho_0 c_0^2$ is a constant capacitance (relevant to the linear case) and 
\begin{equation}
{	C_{\omega}^{'} \simeq \frac{2\beta_{0}}{\rho_{0} c_{0}^{2}}C_{\omega 0}}
	\label{eq:eq5new}.
\end{equation}

We now apply Kirchhoff's voltage and current laws in order to derive the discrete nonlinear dissipative evolution equation for the pressure in the $n$-th cell of the lattice: 
\begin{widetext}
\begin{equation}
%p_{n+1}  -2p_{n}+p_{n-1}
\hat{\delta}^2p_{n} = 
L C_{\omega0}\frac{d^2 p_n}{d t^2}+R_{\omega} C_{\omega0} \frac{d p_n}{d t}+\frac{C_{\omega0}}{C_{m}}p_n  
-\frac{L C_{\omega}^{'}}{2}\frac{d^2 p_{n}^{2}}{d t^2}-\frac{R_{\omega} C_{\omega}^{'}}{2}\frac{d p_{n}^{2}}{d t}-\frac{C_{\omega}^{'}}{C_{m}}p_{n}^{2},
\label{eq:eq6new}
 \end{equation}
\end{widetext}
where $\hat{\delta}^2p_{n} \equiv p_{n+1}  -2p_{n}+p_{n-1}$, and 
$L=L_{\omega}+L_{m}$ (see details in Appendix A). 

Adopting physically relevant parameter values, 
%for this work 
we assume that the distance between the plates is $d=0.01$~m and the clamped elastic plates have a thickness $h=2.78\;10^{-4}$~m, radius $r=0.025$~m, as shown in Fig.~\ref{fig:fig1}(b), and are made of rubber, with $\rho_m=1420$~kg/m$^3$, $E=2.758$ GPa and $\nu =0.34$. Finally, we consider a temperature of $18^{\circ}$~C and the waveguide to be filled by air; thus the specific heat ratio is $\gamma=1.4$, the Prandtl number $Pr=0.71$ and  $\eta=1.84\;10^{-5}$~kg/m/s.

\subsection{Continuum limit}

In order to analytically treat the problem, we focus on the continuum limit of Eq. (\ref{eq:eq6new}), corresponding to $n\rightarrow\infty$ and $d\rightarrow 0$ (with $nd$ being finite). In such a case, the pressure becomes $p_n(t)\rightarrow p(x,t)$, where $x=nd$ is a continuous variable. 
Then, $p_{n \pm 1}$ can be approximated as:
\begin{equation}
{p_{n \pm 1}=p \pm d \frac{\partial p}{\partial x}+\frac{d^2 }{2}\frac{\partial^2 p}{\partial x^2} 
\pm  \frac{d^3}{3!}\frac{\partial^3 p}{\partial x^3}+\frac{d^4}{4!}\frac{\partial^4 p}{\partial x^4}+O(d^5)}, 
\label{eq:eq7new}
\end{equation}
and, accordingly, the operator $\hat{\delta}^2p_{n}$ is approximated as: 
%difference 
%$\hat{\delta}^2p_{n} \equiv p_{n+1}-2p_{n}+p_{n-1}
%$ can be approximated by 
$\hat{\delta}^2 p_n
\approx  d^2 p_{xx} + \frac{d^4}{12} p_{xxxx}$ (subscripts denote partial derivatives). 
Here, having kept terms up to order $O(d^4)$, results in the incorporation of a fourth-order 
dispersion term in the relevant nonlinear partial differential equation (PDE) --see below. 
Including such 
%Keeping the $O(d^4)$ derivative term, the partial differential equation (PDE) also incorporates 
a weak dispersion term, which originates from the periodicity of the elastic plate array (see 
also Ref.~\cite{ref-us}), is necessary in order to capture more accurately the dynamics of  
the system. To this end, neglecting terms of the order $O(d^5)$ and higher, Eq.~(\ref{eq:eq6new}) 
is reduced to the following PDE:
\begin{widetext}
\begin{equation}
	d^2 p_{xx}+ \frac{d^4}{12} p_{xxxx} -LC_{\omega0}p_{tt}-
R_{\omega}	C_{\omega0} p_{t}
-\frac{C_{\omega0}}{C_{m}}p+\frac{1}{2}LC_{\omega}^{'}\left(p^2\right)_{tt}+    
\frac{1}{2}R_{\omega}C_{\omega}^{'}\left(p^2\right)_{t}
+\frac{C_{\omega}^{'}}{C_m}p^2=0
	\label{eq:eq8new}.
\end{equation}  
\end{widetext}
It is also convenient to express our model in dimensionless form; this can be done upon introducing the normalized variables $\tau$ and $\chi$ and the normalized pressure $P$, which are defined as follows: 
$\tau=\omega_B t$ (where $\omega_B=\pi c_0 /d$ is the Bragg frequency), 
$\chi = (\omega_B/c)x$, 
%is the temporal variable in units of $\omega_{B}^{-1}$, where $\omega_B=\pi c_0 /d$ is the Bragg frequency; $\chi$ is the spatial variable in units of $c/\omega_B$, 
where the velocity $c$ is given by
\begin{equation}
	c=\frac{c_0}{\sqrt{1+\alpha}}, \quad 
	\alpha=\frac{h\rho_m}{d\rho_0}
\label{eq:eq9new},
\end{equation}
\noindent
and $p/P_0=\epsilon P$, where $P_0=\rho_0 c_{0}^{2}$ and $0 < \epsilon\ll 1 $ 
is a formal dimensionless small parameter.
Then, Eq. (\ref{eq:eq8new}) is reduced to the following dimensionless form,
\begin{widetext}
\begin{equation}
P_{\tau\tau}-P_{\chi\chi}-\zeta P_{\chi\chi\chi\chi}+\Gamma P_{\tau}
+m^2 P =\epsilon \beta_{0} \left[2 m^2 P^2+\Gamma \left(P^2\right)_{\tau}
+ \left(P^2\right)_{\tau\tau} \right]
\label{eq:eq10new},
\end{equation}
\noindent
where parameters $m^2$, $\zeta$ and $\Gamma$ are given by
\begin{equation}
	m^2=\frac{\alpha}{1+\alpha}\left(\frac{\omega_m}{\omega_B}\right)^2, \quad 
	\zeta=\frac{1}{12}\pi^2(1+\alpha), \quad 
	\Gamma=\frac{R_{\omega}S}{\rho_{0}d\omega_{B}(1+\alpha)}.
	\label{eq:eq11new}
\end{equation}
\end{widetext}

It is interesting to identify various limiting cases of Eq. (\ref{eq:eq10new}). %First, in the lossless linear limit $(R_{\omega}=0$, i.e., $\Gamma=0$ and $\beta_{0}=0$, or $p^2\ll 1)$, and in the absence of plates ($m^2\rightarrow 0$, and without considering  higher order spatial derivatives), Eq. (\ref{eq:eq10new}) is reduced to the 2nd-order linear dispersionless wave equation, $$P_{\tau\tau}-P_{\chi\chi}=0.$$ 
First, in the lossless linear limit $(R_{\omega}=0$, $\Gamma=0$ and $\epsilon\rightarrow0$), in the long-wavelength approximation (without considering  higher-order spatial derivatives, $\zeta\rightarrow 0$), Eq. (\ref{eq:eq10new}) takes the form of the linear Klein--Gordon (KG) equation 
\cite{ref-Remoissenet,ref-Ablowitz}, $$P_{\tau\tau}-P_{\chi\chi}+m^2 P=0,$$ with the parameter $m$ playing the role of mass. If the plates are absent ($m^2\rightarrow 0$) the Klein--Gordon equation is reduced 
to the 2nd-order linear wave equation. Another interesting limit of Eq. (\ref{eq:eq10new}) 
%consists of assuming 
corresponds to $m^2\rightarrow 0$, $\Gamma=0$ and $\zeta\rightarrow 0$, which leads to the well-known Westervelt equation, $$P_{\tau\tau}-P_{\chi\chi}-\epsilon\beta_{0}\left(P^2\right)_{\tau\tau}=0,$$ which is a common nonlinear model describing 1D acoustic wave propagation \cite{ref-Hamilton}. 

\subsection{Linear limit}

We now consider the linear limit ($\epsilon\rightarrow 0$) of Eq.~(\ref{eq:eq10new}), 
and assume propagation of plane waves of the form $P\propto \exp[i(k\chi-\omega \tau)]$, 
to obtain the following dispersion relation
\begin{equation}
	D(\omega,k)=(-\omega^2+k^2-\zeta k^4+m^2)-i\Gamma\omega=0
	\label{eq:eq12new}.
\end{equation}

\begin{figure}[tbp]
 \centering
\includegraphics[width=8.5cm]{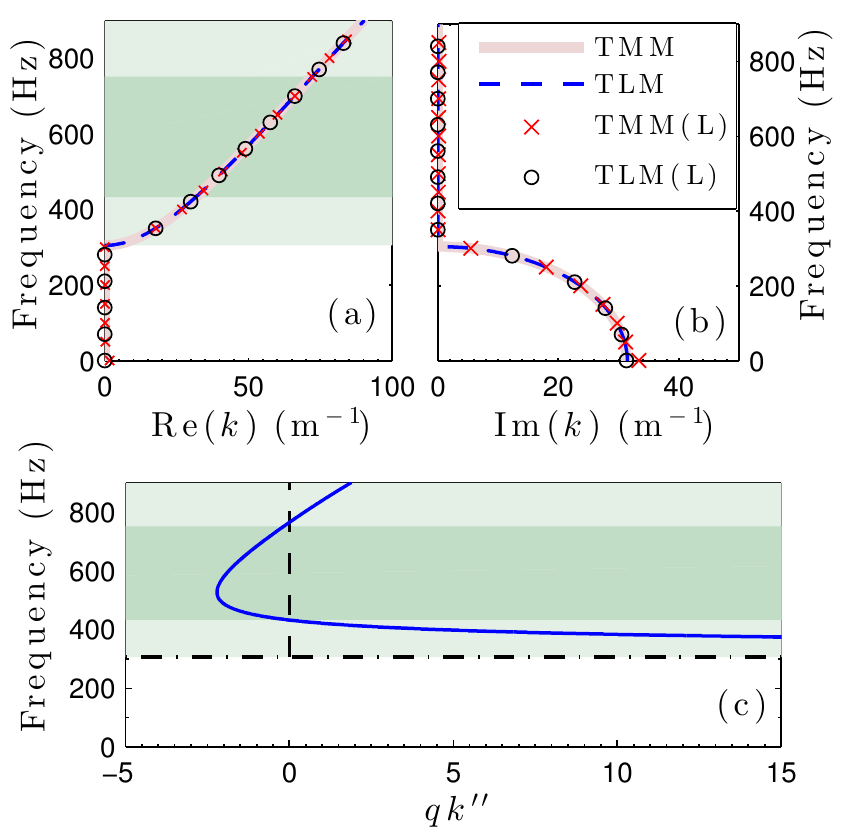} 
  \caption{(Color online) (a) and (b) respectively show the real and imaginary 
  parts of the complex dispersion relation. 
  %of system analyzed in this work. 
  Black circles (Red crosses) show the results from the TL (TMM) approach from 
  Eq.~(\ref{eq:eq13new}) (Eq.~(\ref{eq:eq14new})). Blue dashed (light pink continuous) line 
  shows the lossless dispersion relation obtained from the TL (TMM) approach 
  from the lossless limit of Eq.~(\ref{eq:eq13new}) (Eq.~\ref{eq:eq14new}). 
  (c) The frequency dependence of $qk''$, the product of the dispersion and nonlinearity 
  coefficients of the NLS equation. Light (dark) green region corresponds to 
  the focusing (defocusing) case, with $qk''>0$ ($qk''<0$).
%  (focusing case)   and dark green region to $qk''<0$ (defocusing case).
  } 
  	\label{fig:fig2}
\end{figure} 

%For $\zeta=0$ and $\Gamma=0$, this is the familiar lossless dispersion relation of the linear Klein--Gordon model. It is clear that 
Equation (\ref{eq:eq12new}) suggests the existence of a gap at low frequencies, i.e., for $0\leq\omega<m$, with the cut-off frequency defined by the parameter $m$ 
(as is common in KG--type models \cite{ref-Remoissenet,ref-Ablowitz}). 
For $m<\omega<\omega_B$, there exists a propagating band, with the dispersion curve $\omega(k)$ having the form of hyperbola, which asymptotes [according to Eq.~(\ref{eq:eq12new})] to unity, representing the normalized velocity associated with the linear wave equation $P_{\tau\tau}-P_{\chi\chi}=0$ 
mentioned above. The term $\zeta k^4$ appears to lead to 
instabilities for large values of $k$. However, both Eqs.~(\ref{eq:eq10new}) and (\ref{eq:eq12new}) 
are used in our analysis only in the long-wavelength limit where $k$ is sufficiently small. The term  $i\Gamma\omega$ accounts for the viscothermal losses. 

Since all quantities in the dispersion relation are dimensionless, it is also relevant 
to express Eq.~(\ref{eq:eq12new}) in physical units. In particular, taking into account 
that the frequency $\omega_{ph}$ and wavenumber $k_{ph}$ in physical units are connected 
with their dimensionless counterparts through $\omega=\omega_{ph}/\omega_B$ and 
$k=\frac{k_{ph} c}{\omega_B}$, we can express Eq.~(\ref{eq:eq12new}) in the following form:
\begin{equation}
	-\omega_{ph}^{2}+k_{ph}^{2}c^2-\zeta\frac{k_{ph}^{4}c^4}{\omega_{B}^{2}} +{m}^{2}\omega_{B}^{2}-i\Gamma\omega_{ph}\omega_{B}=0.
	\label{eq:eq13new}
\end{equation}
The real and imaginary parts of the dispersion relation are respectively plotted in 
Figs.~\ref{fig:fig2}(a) and \ref{fig:fig2}(b). We observe that there is almost 
no difference between the lossy dispersion relation [Eq. (\ref{eq:eq13new})] and the lossless one 
[Eq.~(\ref{eq:eq13new}) with $\Gamma=0$], since the losses are sufficiently small (see below). 
The dispersion relation features the band gap from $0$~Hz to $\left(m \frac{\omega_B}{2\pi} \right)$~Hz 
due to the combined effect of the resonance of the plate and of the geometry of the system. 
The upper limit of the band gap is found to be sufficiently smaller than the Bragg band frequency 
$f_B=c_0/2d=17163$~Hz, with $c_0=343.26$~m/s. 
We have compared this analytical dispersion relation with the one obtained via the 
transfer matrix method (TMM) \cite{ref-Bradley}. 
Solid (light pink) lines and (red) crosses in the Figs.~\ref{fig:fig2}(a) and \ref{fig:fig2}(b) 
show the respective results, as obtained using the TMM from the following relation \cite{ref-Bradley}: 
    \begin{equation}
    \cos(k_{ph} d)=\cos( k d)+i \frac{Z_m}{2 Z_c}\sin(k d),
    \label{eq:eq14new}
    \end{equation}
%\noindent    
%where 
%%\begin{equation}
%%k=\frac{\omega_{ph}}{c_0}\left( 1+\frac{1-j}{s}(1+(\gamma-1)/B) \right),
%%\end{equation}
%%\begin{equation}
%% Z=\frac{\rho_0 c_0}{S}\left( 1+\frac{1-j}{s}(1-(\gamma-1)/B) \right),
%%\end{equation}
%%\noindent
where $Z_m=i\left(\omega_{ph} L_m -1/\omega_{ph} C_m\right)$ is the impedance of the plate 
\cite{ref-Theocharis}, and $Z_c$ is given by Eq.~(\ref{zc}). For the lossless case 
[solid (pink) lines in Figs.~\ref{fig:fig2}(a) 
and \ref{fig:fig2}(b)], the wavenumber and the acoustic characteristic impedance 
of the waveguide 
%turns 
reduce to $k=\omega_{ph}/c_0$ and $Z_c=\rho_0 c_0/S$ respectively. 
Comparing the dispersion relation obtained by using TMM, with the one resulting from the continuum approximation, we find an excellent agreement between these two in the regime of low frequencies.

\section{Envelope solitons}\label{sec3}

%In this Section, we will derive NLS envelope solitons (bright and gap ones) in the case of the lossless clamped elastic plate lattice model. Our methodology relies on the use of the multiple scales perturbation method, by means of which Eq. (\ref{eq:eq10new}) is reduced to an effective NLS equation. The effect of losses will be studied below (cf; Section \ref{sec4}).
In this Section, we apply the multiple scales perturbation method to reduce Eq.~(\ref{eq:eq10new}) to an effective NLS equation. This way, we derive approximate analytical envelope soliton solutions  
%We then study the dynamics of envelope solitary solutions 
of the original lattice system, and study their dynamics --by means of direct numerical simulations 
and soliton perturbation theory-- 
%which corresponds to the effective NLS solitons
in the absence and presence of viscothermal losses.

\subsection{Bright solitons: propagating solitary waves}\label{sec3A}

 \begin{figure*}
 \centering
 \includegraphics[width=16cm]{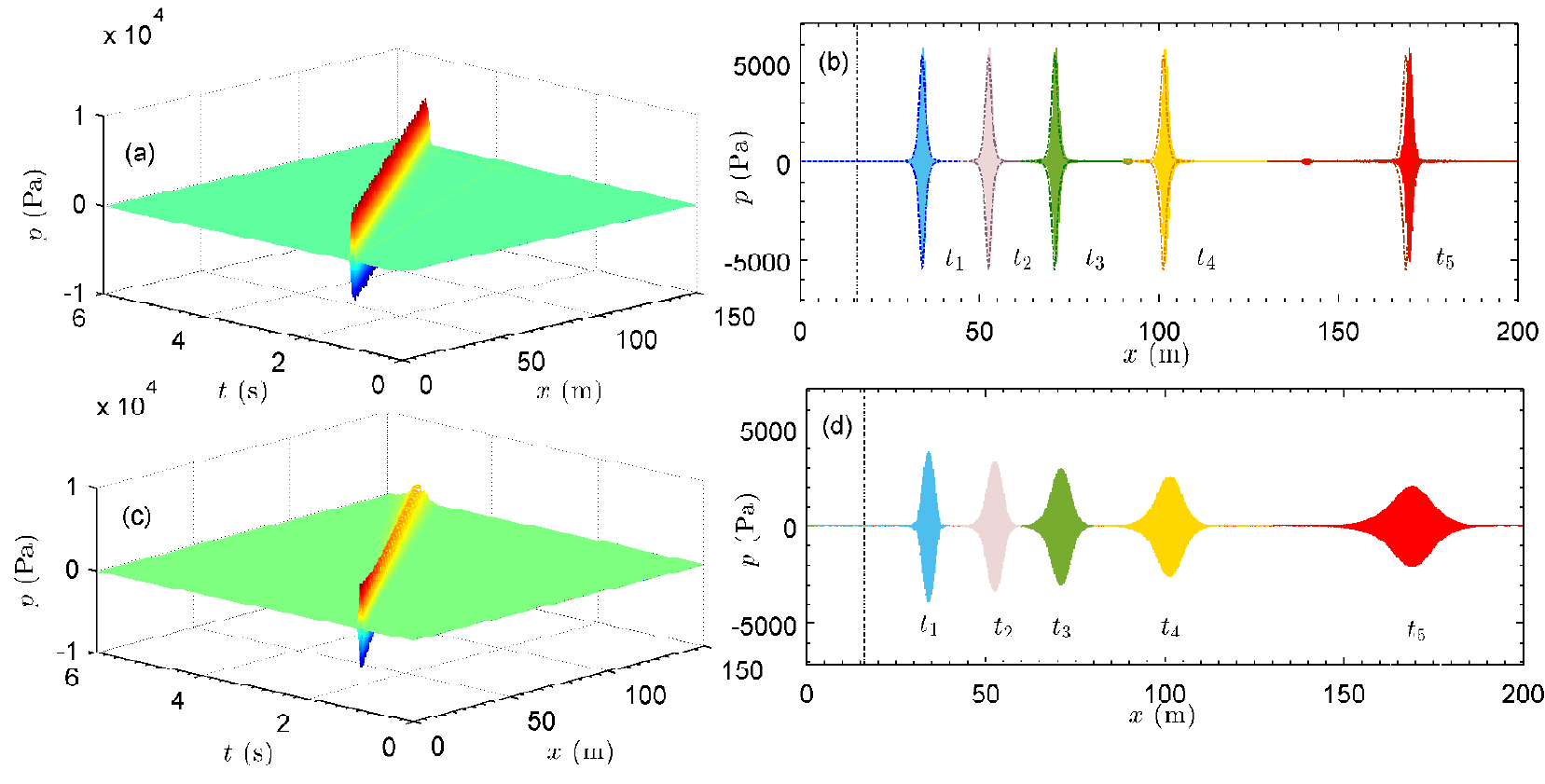} 
  \caption{ (Color online) (a) 3D plot depicting the evolution of a bright soliton of the form of Eq. (\ref{eq:eq26new}), obtained by numerically integrating the lossless version of Eq. (\ref{eq:eq6new}) ($R_{\omega}=0$) with $\epsilon=0.018$ ($2\epsilon\eta P_0=5471$ Pa), $f=369$ Hz. (b) Numerical spatial profiles of the bright soliton measured at $t_1=2$ s (light blue line), $t_2=2.5$ s (light pink line), $t_3=3$ s (light green line), $t_4=4$ s (light yellow line), and $t_5=5.7$ s (red line). Blue dashed line, pink dashed line, green dashed line, yellow dashed line, and dark red dashed line stand for the analytical envelope results of Eq. (\ref{eq:eq26new}) at $t_1$, $t_2$, $t_3$, $t_4$ and $t_5$ respectively. Black dash-dotted line stand for the nonlinear length $L_{NL}$ and the dispersion length $L_D$, where $L_{NL}=L_{D}=16$ m; (c) Turn off the nonlinearity effect, 3D plot depicting the dispersive effect numerically obtained. (d) Numerical spatial profile of dispersive effect measured at $t_1$ (light blue line), $t_2$ (light pink line), $t_3$ (light green line), $t_4$ (light yellow line), and $t_5$ (red line).} 
\label{fig:fig3}
\end{figure*} 

We start our analysis by introducing the slow variables,
\begin{equation}
\chi_n=\epsilon^n \chi, \quad \tau_n=\epsilon^n \tau, \quad n=0,1,2,\ldots,
\label{eq:eq15new}
\end{equation}
\noindent
 and express $P$ as an asymptotic series in $\epsilon$, namely: 
 \begin{equation}
P=p_0+\epsilon p_1+\epsilon^2 p_2+\ldots.
\label{eq:eq16new}
\end{equation}
Then, substituting the above into Eq.~(\ref{eq:eq10new}) we obtain a hierarchy of equations 
at various orders in $\epsilon$ (see Appendix B). This way, and assuming that the 
losses are sufficiently small, 
%considering that we are in the case of small enough losses, 
namely $\Gamma \rightarrow \epsilon^2\Gamma$, we obtain the following results.
%, which means that below we will treat terms $\propto \Gamma$ as small perturbations.

First, at the leading order, i.e. at $O(\epsilon^0)$, we find that $p_0$ satisfies a linear 
wave equation [cf. Eq.~(B1) in Appendix~B], and thus  
%the plane wave solution 
is of the form:
\begin{equation}
p_0=A(\chi_1,\chi_2,\cdots,\tau_1,\tau_2,\cdots) 
\exp(i \theta)+{\rm c.c.},
\label{eq:eq17new}
\end{equation}
where $A$ is an unknown envelope function, 
%of $p_0$, the phase 
$\theta=k\chi_0-\omega \tau_0$, with the wavenumber $k$ and frequency $\omega$ satisfying 
the dispersion relation~(\ref{eq:eq12new}) (c.c. denotes complex conjugate). 

Next, 
%we consider the equation 
at the order $O(\epsilon^1)$, we obtain an equation whose 
%Its 
solvability condition requires that the secular part [i.e., the term $\propto \exp(i \theta)$]
vanishes. 
This yields the following equation,
\begin{equation}
\left(k' \frac{\partial }{\partial \tau_1}+\frac{\partial }{\partial \chi_1}\right) A(\chi_1,\chi_2,\cdots,\tau_1,\tau_2,\cdots) =0,
\label{eq:eq18new}
\end{equation}
where the inverse group velocity $k' \equiv \partial k/\partial \omega = 1/v_g$ is given by
\begin{equation}
k' 
%\equiv \frac{\partial k}{\partial\omega}
=\frac{\omega}{k-2\zeta k^3}.
\label{eq:eq19new}
\end{equation}
Equation (\ref{eq:eq18new}) is satisfied as long as $A$ depends on the variables $\chi_1$ and $\tau_1$ 
through the traveling-wave coordinate $\tilde{\tau}_{1}=\tau_{1}-k'\chi_{1}$ 
(i.e., $A$ travels with the group velocity), namely $A(\chi_1, \tau_1,\chi_2, \tau_2, \cdots)=A(\tilde{\tau}_1,\chi_2, \tau_2, \cdots)$. Additionally, at the same order, 
we obtain the form of the field $p_1$, namely,
\begin{equation}
p_1=2\beta_0 \frac{m^2-2\omega^2}{D(2\omega,2k)}A^2 e^{2i\theta}+B  e^{i\theta}+4 \beta_0 \left|A\right|^2 +{\rm c.c.},
\label{eq:eq20new}
\end{equation}
where 
%$4 \beta_0 \left|A\right|^2$ is a DC term and 
$B$ is an unknown function that can be found at a higher-order approximation.

Finally, employing the non-secularity condition at $O(\epsilon^2)$, yields 
%a lossy NLS equation 
the following PDE for the envelope function $A$,
\begin{equation}
i\frac{\partial A}{\partial \chi_{2}}-\frac{1}{2} k''\frac{\partial^2 A}{\partial \tilde{\tau}_{1}^{2} }-q \left|A\right|^2 A=-i \Lambda A,
\label{eq:eq21new}
\end{equation}
which is a NLS equation incorporating a linear loss term. The 
%where the 
dispersion, nonlinearity and dissipation coefficients are respectively given by:
\begin{equation}
k'' \equiv \frac{\partial^2 k}{\partial \omega^2}=\frac{1-k'^{2}+6\zeta k^2 k'^{2}}{k-2 \zeta k^3},
\label{eq:eq22new}
\end{equation}
\begin{equation}
q(\omega,k)=\beta_0^2  \frac{2(2m^2-\omega^2)(m^2-2\omega^2)}{3(m^2+4\zeta k^4)(k-2 \zeta k^3)}-\beta_0^2 \frac{4(2m^2-\omega^2)}{(k-2 \zeta k^3)},
\label{eq:eq23new}
\end{equation}
\begin{equation}
\Lambda=\frac{\omega\Gamma}{2 (k_{r}-2 \zeta k_{r}^3)}.
\end{equation}

The sign of the product $\sigma \equiv {\rm sgn}(q k'')$ determines the nature of 
the NLS equation and its solutions \cite{ref-Remoissenet,ref-Ablowitz}. 
In particular, in the case 
%with 
$\sigma=+1$ ($\sigma=-1$) the NLS 
is called focusing (defocusing) and supports 
%the system is characterized by an anomalous (normal) dispersion allowing for 
bright (dark) soliton solutions. Bright solitons are localized waves with 
vanishing tails towards infinity, while dark solitons are density dips, with a phase jump 
across the density minimum, on top of a non-vaninishing continuous wave background. 
%extended constant amplitude waves with a localized dip.
%(solitary waves localized in space, and having spatial attenuation towards infinity). 
%The other case with $\sigma=-1$ is known as defocusing, the system is characterized by a normal dispersion, and in this case the NLS does not admit solitons that vanish at infinity, but supports dark soliton solutions.%, solitary waves having constant amplitude at infinity, and a local spatial dip in amplitude.
Figure~\ref{fig:fig2}(c) shows the frequency dependence of the product $q k^{''}$ for the system. We observe three different regimes: 
focusing regime ($\sigma=+1$) at low frequencies (light green region), defocusing regime ($\sigma=-1$) 
at 
%medium 
intermediate frequencies (dark green region), and another focusing regime ($\sigma=+1$) 
at high frequencies (light green region). Below we focus on the case of the focusing NLS equation 
and study propagating bright solitons and stationary gap solitons that are supported in this case. 

The dispersion length, $L_D$, and the nonlinearity length, $L_{NL}$, 
provide the length scales over which dispersive or nonlinear effects 
become important for pulse evolution. For 
%the case of 
solitons, where 
%in which 
the nonlinearity and dispersion effects should be perfectly balanced, $L_D\simeq L_{NL}$ (see Appendix C for details).  %\textcolor{blue}{ 
For frequencies larger than $435$ Hz, the dispersion is very weak leading 
(e.g., for $\epsilon=0.018$ and $f=435$~Hz) to a dispersion length of the order of 
$L_D=450$~m.
%$...$ meters.
%Thus, in this work 
Here, we focus on the low frequency region (light green region from $305.7$ Hz to $432.3$ Hz) described by an effective focusing NLS with linear loss, in order to study 
%where the dispersion is strong and 
the combined effect of (a relatively strong) dispersion and nonlinearity. % can be studied. 
%}
%It is found that the dispersion length \sout{in the top two regions} \textcolor{red}{for high frequency} is huge due to the weak dispersion. Thus, we focus on the %region at lower frequencies (light green region from $305.7$ Hz to $432.3$ Hz) where the dispersion is strong and the combined effect of dispersion and %nonlinearity can be studied.

\subsubsection{Bright solitons in the absence of losses}
%Lossless bright solitary waves}
In the absence of losses ($\Gamma=0$), the analytical bright soliton solution for the envelope 
function $A$ is of the form, 
\begin{equation}
A=\eta \textrm{sech}\left(\eta \sqrt{\left| \frac{q}{k^{''}}\right| }\tilde{\tau}_1\right) 
\exp\left(-i \frac{q \eta^2}{2} \chi_2\right),
\label{eq:eq24new}
\end{equation}
where $\eta$ is a free parameter setting the soliton amplitude. 
The corresponding approximate solution of Eq.~(\ref{eq:eq10new}) 
is expressed, as a function of parameters $\chi$ and $\tau$, 
%in terms of coordinates $\chi$ and $\tau$ is written 
as follows:
\begin{widetext}
\begin{equation}
P(\chi, \tau) \approx  2 \eta \textrm{sech}\left[ \epsilon \eta \sqrt{\left| \frac{q}{k''}\right| } \left(\tau-k' \chi \right)  \right]  \cos\left(\omega\tau-k\chi-\frac{q \epsilon^2\eta^2}{2}\chi \right).
\label{eq:eq25new}
\end{equation}
Futhermore, in the original space and time coordinates, $x$ and $t$, 
the approximate envelope soliton solution for the pressure $p$ reads:
\begin{equation}
\frac{p(x,t)}{P_0}\approx  2 \epsilon \eta \textrm{sech}\left[ \epsilon \eta \sqrt{\left| 
\frac{q}{k''}\right| } \omega_B \left(t-\frac{k'\sqrt{1+\alpha}}{c_0 }x \right) \right] 
\cos\left(\omega \omega_B t-k\omega_B\frac{\sqrt{1+\alpha}}{c_0}x-\frac{q \epsilon^2\eta^2}{2}\omega_B\frac{\sqrt{1+\alpha}}{c_0}x \right).
\label{eq:eq26new}
\end{equation}
\end{widetext}
This bright soliton is characterized by an amplitude $2 \epsilon \eta P_0$ and 
a width $(\epsilon\eta\sqrt{|\frac{q}{k''}|})^{-1}$. 
In addition, its velocity is given by the group velocity $c_0/(k'\sqrt{1+\alpha})$ at the carrier frequency and in contrast with soliton solutions of other nonlinear dispersive wave equation 
other soliton solutions, like for instance the soliton of 
e.g., the Korteweg-de Vries (KdV) equation 
%KdV soliton 
\cite{ref-Ablowitz} is independent of its amplitude.

 \begin{figure*}
 \centering
 \includegraphics[width=16cm]{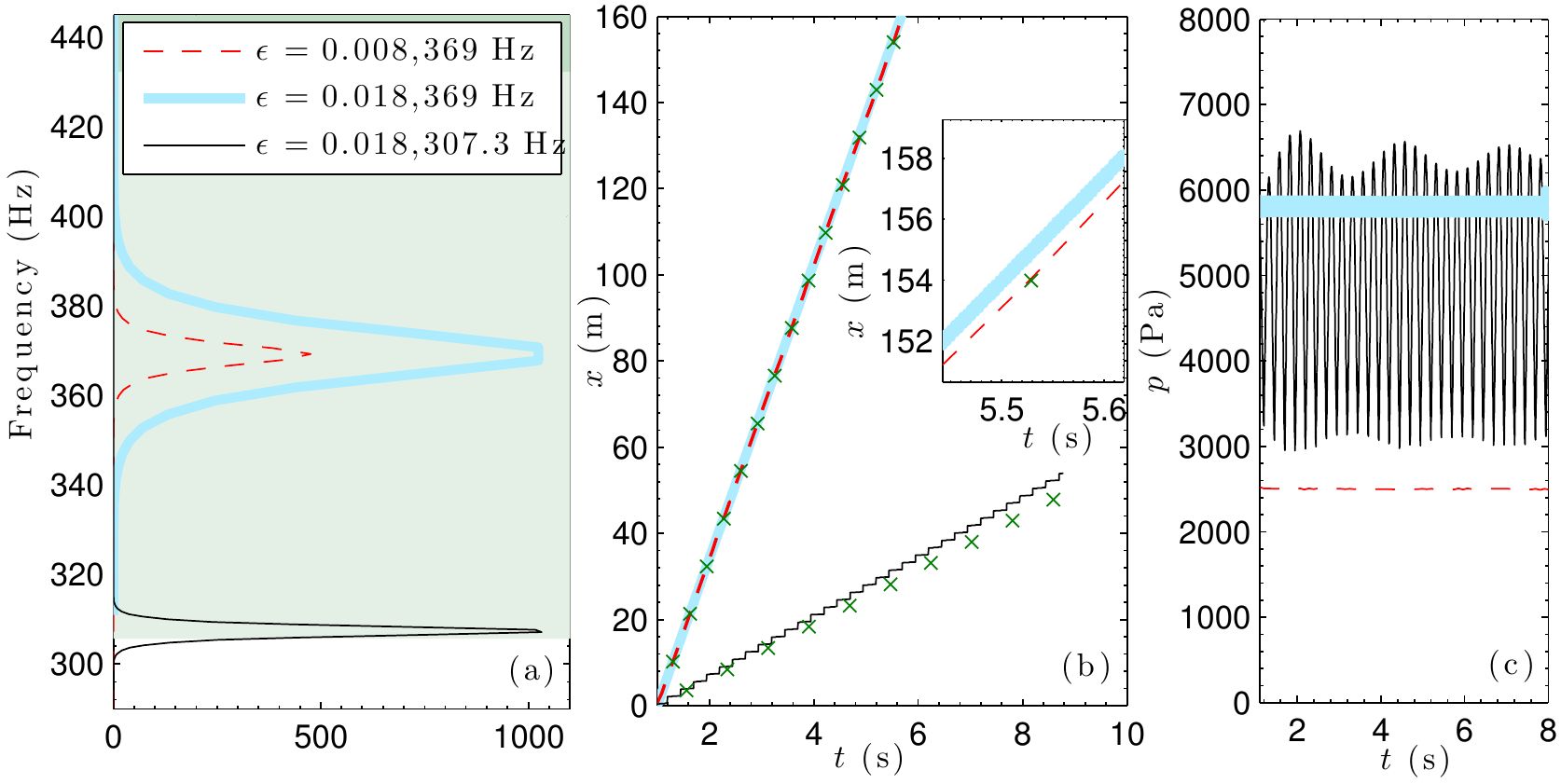} 
  \caption{ (Color online)(a) Spectra of the different drivers, of the form of Eq. (\ref{eq:eq26new}), introduced at $x=0$: $\epsilon=0.008$ ($2\epsilon\eta P_0=2431$ Pa) at $369$ Hz (red dashed line); $\epsilon=0.018$ ($2\epsilon\eta P_0=5471$ Pa) at $369$ Hz (thick light blue continuous line) and $\epsilon=0.018$ at $307.3$ Hz (thin black continuous line). (b) Space-time diagrams of the different wave packet generated from the different drivers in (a). Symbols stand for the analytical space-time diagrams at $369$ Hz and $307.3$ Hz. The slopes of the lines depict the corresponding group velocities. (c) Numerical time evolutions of the maximum pressure value of the solitons for the different drivers.} 
  	\label{fig:fig4}
\end{figure*} 

Let us now proceed by studying numerically the evolution of the 
approximate soliton solution of Eq.~(\ref{eq:eq26new}), 
in the framework of the fully discrete model of Eq.~(\ref{eq:eq6new}). We start with the lossless case
($R_{\omega}=0$) and a driver of the form given by Eq.~(\ref{eq:eq26new}) at $x=0$.
% by means of the Runge-Kutta method. For each simulation, we ensure the validity of the Courant--Friedrichs--Lewy (CFL) condition, $c \frac{dt}{dx} \leq 1$, where $c$ is the phase velocity, $dt$ and $dx$ are the time step and length interval, respectively.
%We also pay attention to the length of the system, which should be long enough to avoid reflections of the propagating wave. 
We use the parameter values $\epsilon=0.018$ ($2\epsilon\eta P_0=5471$~Pa) and $f=369$~Hz. 
The results of the simulations are shown in Figs.~\ref{fig:fig3}(a) and \ref{fig:fig3}(b).
We observe that the input envelope wave propagates with a constant amplitude 
and width as shown in the spatio-temporal evolution in Fig.~\ref{fig:fig3}(a). 
%Indeed,the wave propagates with no distortion after a distance corresponding to the nonlinear length (equal to the dispersion length shown by the vertical dotted line in Fig. \ref{fig:fig3}(b)). Due to the perfect balance between the nonlinear and dispersive effect the wave packet propagates keeping the same amplitude and width as shown in the space/time map in Fig. \ref{fig:fig3}(a). 
The direct comparison of analytics and simulations is shown in Fig.~\ref{fig:fig3}(b). 
Here, the envelope soliton solution 
%calculated by the analytical expression 
of Eq.~(\ref{eq:eq26new}), is compared at five different instants with the numerical 
results for the discrete wave equation showing a very good agreement. 
Thus, we confirm that the NLS approximation is able to capture the propagation 
of envelope solitons of the discrete model~(\ref{eq:eq6new}). 
%Small deviations appears only at large time $t_5$ ($5.7$ s).
To emphasize the effect of the counterbalance of dispersion by nonlinearity, 
we also show the evolution of the same envelope function when the nonlinearity is turned off. 
As shown in Figs.~\ref{fig:fig3}(c) and \ref{fig:fig3}(d), the initial wavepacket spreads 
as it propagates due to dispersion.

Next, we study the validity of the multiple-scales perturbation theory and the properties of the corresponding bright solitons. To do so, we study three different solutions: two at the same carrier frequency $f=369$ Hz with different amplitudes, $\epsilon=0.008$ ($2\epsilon\eta P_0=2431$ Pa), and $\epsilon=0.018$ ($2\epsilon\eta P_0=5471$ Pa) and one of amplitude  $\epsilon=0.018$ ($2\epsilon\eta P_0=5471$ Pa) and carrier frequency $f=307.3$ Hz. The respective spectra of these solitons are depicted in Fig.~\ref{fig:fig4}(a). Note that, for the last case, part of the spectrum of the soliton lies inside the gap.
Starting with the two soliton solutions at the same carrier frequency but different amplitudes, we expect them to propagate with the same velocity, i.e., the group velocity. In Fig.~\ref{fig:fig4}(b), 
the dashed red and solid cyan lines show the 
the position of the maximum of the numerical solution as a function of time, for $\epsilon=0.008$ and $\epsilon=0.018$, respectively. Green crosses depicts the analytical group velocity. 
Both solutions appear to follow with a very good agreement with 
the analytical prediction. In addition, as shown in  Fig.~\ref{fig:fig4}(c), 
these solutions propagate with constant amplitude.
However, as seen in 
%looking %carefully at 
the inset of Fig.~\ref{fig:fig4}(b), there is a small discrepancy 
%is observed 
in the velocity of the envelope solutions 
%with the 
of larger amplitude. This indicates a deviation from the effective NLS description for large amplitudes, 
which is naturally expected due to the perturbative nature of our analytical approach.
Note, that this small deviation is also depicted in Fig.~\ref{fig:fig3}(b) for the last time instant.

The third case corresponds to the solution whose part of its spectrum lies in the gap, 
for $\epsilon=0.018$ and $f=307.3$ Hz.
Here, we observe the propagation of a breathing solitary solution. The respective
long-lived, weakly damped periodic oscillations of the soliton amplitude are depicted in 
Fig.~\ref{fig:fig4}(c).
As it has been discussed \cite{ref-internalmode1,ref-internalmode2}, 
this behavior may be associated to the birth of an internal mode of the soliton. 
We also observe a small deviation between the numerical group velocity 
and the corresponding analytical one, as shown in Fig.~\ref{fig:fig4}(b).

%In this case part of the frequencies of the driver is outside the focusing region, as shown in Fig. \ref{fig:fig4}(a). The numerical group velocity, shown in Fig. \ref{fig:fig4}(b), is not equal to the corresponding analytical one: the numerical group velocity is slightly bigger than the analytical one. Moreover, as shown in Fig. \ref{fig:fig4}(c), the maximum amplitude of the soliton is no longer constant during the propagation. It oscillates and we can also observe the beatings due to the combined effect of bright solitons and the internal mode, which is responsible for long-lived, weakly damped periodic oscillations of the soliton amplitude observed in many numerical simulations  \cite{ref-internalmode1, ref-internalmode2}.

\subsubsection{Bright solitary waves in the presence of losses}

\begin{figure}[t]
 \centering
 \includegraphics[width=8.5cm]{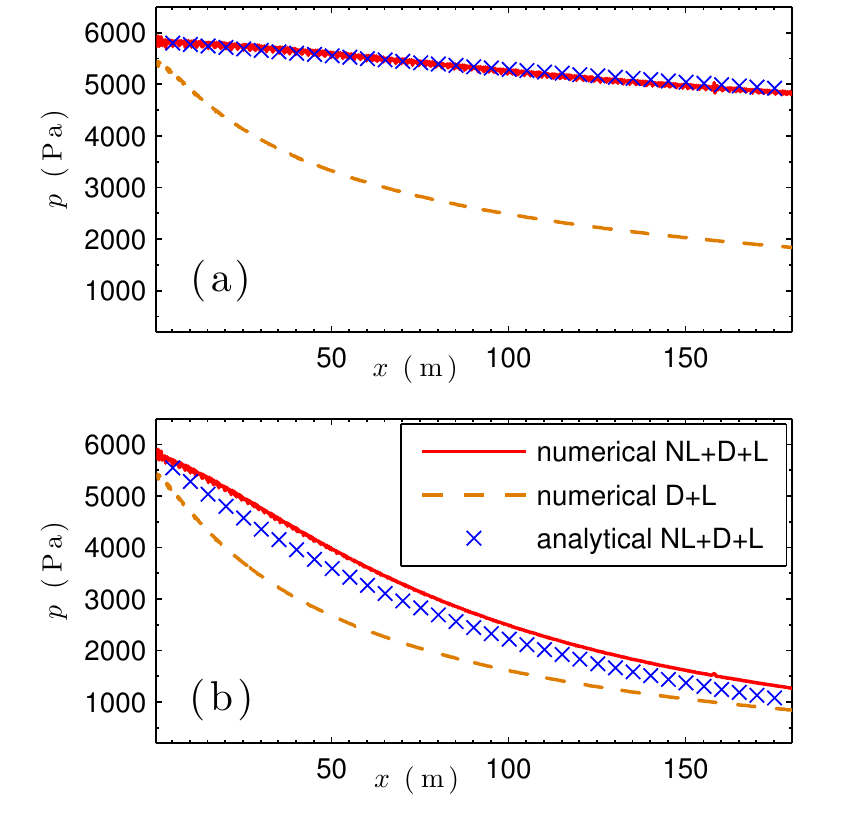} 
  \caption{(Color online) Effect of viscothermal losses on traveling bright solitons. Evolution of the maximum pressure in time for the lossy bright soliton (continuous red line for numerical results and blue crosses for the analytical ones) and for linear lossy dispersive wave (dashed yellow line for numerical results). The driver corresponds to $\epsilon=0.018$ ($2\epsilon\eta P_0=5471$ Pa) and $f=369$ Hz. (a) Propagation in a weakly lossy medium where $R_{\omega}=6.8$ Ohm; (b) Propagation in a real lossy medium where $R_{\omega}=68.04$ Ohm.} 
  	\label{fig:fig8}
\end{figure} 

Having established the validity of the NLS solitons in the lossless version of the 
%fully, lossless 
discrete model~(\ref{eq:eq6new}), 
we proceed by studying the evolution 
%robustness 
of the envelope solitons under the presence of the viscothermal losses.
We numerically integrate the nonlinear lattice model with $R_{\omega}=6.8$ Ohm 
and with $R_{\omega}=68.04$ Ohm, using a driver corresponding to the soliton shown 
in Fig.~\ref{fig:fig4} with parameters $\epsilon=0.018$, and $f=369$ Hz.

As shown in Fig.~\ref{fig:fig8}(a), for the small resistor of $R_{\omega}=6.8$ Ohm, 
%in the presence of weak losses ($R_{\omega}=6.8$ Ohm), 
the amplitude of the soliton is found to be weakly attenuated. 
In contrast, in the linear dispersive case 
(see dashed orange line) the combined effect of dispersion and losses strongly 
attenuates the wave packet. 
Let us next consider the case of the large resistance, $R_{\omega}=68.4$ Ohm, 
corresponding to the viscothermal losses at $f=369$ Hz, assuming 
%considering 
an air-filled waveguide at $18^{\circ}$ C. In this case, 
as shown in Fig.~\ref{fig:fig8}(b) 
%for the case of the large resistance, $R_{\omega}=68.4$ Ohm, 
%presence of strong losses ($R_{\omega}=68.4$ Ohm), as shown in Fig. \ref{fig:fig8}(b), 
the effect of losses on the soliton amplitude is (naturally) more pronounced, 
but still less than the case without considering the nonlinearity 
%effect 
[dashed yellow line in Fig.~\ref{fig:fig8}(b)]. 
%; this is due to the fact that 
% because nonlinearity can partly balance the dispersion effect. 
%Note that the stronger dissipation considered in the second case correspond to the viscothermal 
%losses at $f=369$ Hz, assuming 
%%considering 
%an air-filled waveguide at $18^{\circ}$ C. 
Here, we also mention that the above findings are valid for the particular 
(physically relevant) scenarios discussed above. Indeed, generally, 
since --as discussed above-- dispersion, nonlinearity and dissipation set 
pertinent scales, it is exactly this scale competition that defines the 
nature of the dynamics.

%%%%%%%%%%%%%%%%%%%%%%%%%%%%%%%%%%%%%%%%%%%%%%%%%%%%%%%%%%%%%%%%%%%%%%%%%
%%%%%%%%%%%%%%%%%%%%%%%%%%%%%%%%%%%%%%%%%%%%%%%%%%%%%%%%%%%%%%%%%%%%%%%%%

Losses have been considered weak in the multiple-scale perturbation method, leading to the 
effective NLS~(\ref{eq:eq21new}) with the linear loss. Furthermore, as long as 
the parameter $\Lambda$ is 
small enough, it is possible 
%This allows us 
to analytically study the role of such a dissipation 
%influence in the NLS 
on the soliton dynamics. Indeed, according to  
%by considering  them as a weak perturbation ($\Lambda \ll 1$) and employing the 
soliton perturbation theory  
%As it is shown in Ref. 
(see, e.g., Ref.~[\onlinecite{ref-NonlinearFiberOptics}]), the linear loss 
%the losses do not influence 
does not affect the velocity of the soliton but its amplitude $\eta$ becomes 
a decaying function of time. The evolution of $\eta$, can be determined by the 
evolution of the norm, and it is straightforward to find that it is described as follows:
%given by the following expression:
\begin{equation}
\eta(\chi_2)=\eta(0)\exp(-2\Lambda \chi_2).
\end{equation}
\noindent
Thus, in terms of the original coordinates, the amplitude of the bright soliton decreases 
exponentially according to:
\begin{equation}
\eta(x)=\eta(0) \exp\left(-2\Lambda \epsilon^2 \omega_B \frac{\sqrt{1+\alpha}}{c_0 }x\right).
\end{equation}
\noindent
This analytical result is denoted in Fig.~\ref{fig:fig8} by crosses. For the case 
of $R=6.8$ Ohm --Fig.~\ref{fig:fig8}(a)-- the agreement between numerical simulations 
and soliton perturbation theory is excellent. For the case of $R=68.04$ Ohm 
--Fig.~\ref{fig:fig8}(b)-- the analytical result describes fairly well the 
amplitude attenuation observed in simulations. For both cases, we also confirm in 
%numerics
the simulations that the envelope solutions propagate with a constant velocity equal to $v_{g}$. 
We can thus conclude that even in the presence of realistic viscothermal losses, 
the system supports envelope solitary waves that are described, in a very good approximation, 
by the effective NLS~(\ref{eq:eq21new}) with the linear loss.

\subsection{Gap solitons: stationary solitary waves}\label{sec4}

%In the Section \ref{sec3A}, we introduced the bright soliton with an envelope moving at velocity $v_g$. We note that for $v_g=0$, the envelope soliton does not move and it oscillates at one frequency in the band gap of the system, this situation corresponds to gap solitons. We will study this case in detail in the current Section. 
While in Sec.~\ref{sec3A} we introduced the traveling bright soliton
propagating with group velocity $v_g$, 
%with an envelope moving at velocity $v_g$. %For the particular case with $v_g=0$, the envelope soliton does not move and it oscillates at one frequency in the band gap of the system, this situation corresponds to gap solitons. 
now we will study stationary (i.e., non-traveling) localized waveforms 
%solutions 
oscillating at a frequency in the band gap of the system; these structures are 
called gap solitons. 
    
%Gap solitons are particular cases of Eq. (\ref{eq:eq26new}) with $k=0$ and $\omega=m$, but it is impossible to set directly $k=0$ and $\omega=m$ in Eq. (\ref{eq:eq26new}), ($k'=\infty$). 
In order to identify such solitons, which evolve in time rather than space, 
we need to derive a variant of the NLS model with the evolution variable being the time.
To do so, returning back to our perturbation scheme, in the solvability condition 
of the equation at the order $O(\epsilon^1)$, we use the variable 
%a new system of coordinates, 
$\xi_{1}=\chi_{1}-v_g \tau_{1}$. This way, we obtain:
\begin{equation}
\left( \frac{\partial }{\partial \tau_1}+v_g \frac{\partial }{\partial \chi_1}\right) A(\chi_1,\chi_2,\cdots,\tau_1,\tau_2,\cdots) =0,
\label{eq:eq27new}
\end{equation}
%where the group velocity $v_g$ is given by
%\begin{equation}
%v_g = \frac{\partial \omega}{\partial k}=\frac{k-2\zeta k^3}{\omega}.
%\label{eq:eq28new}
%\end{equation}
which is satisfied as long as $A$ depends on the variables $\chi_1$ and $\tau_1$ through the traveling-wave coordinate $\tilde{\xi}_{1}$, namely $A(\chi_1, \tau_1,\chi_2, \tau_2, \cdots)=A(\xi_1,\chi_2, \tau_2, \cdots)$ 
%. In this new coordinate, we note that 
[in this case, $p_1$ is again given by 
%the expression for $p_1$ is the same as that in the 
Eq.~(\ref{eq:eq20new})]. %In this case, 
Then, the non-secularity condition at $O(\epsilon^2)$ leads to the following NLS equation,
\begin{equation}
i\frac{\partial A}{\partial T_{2}}-\frac{1}{2} v_{g}^{3} k^{''}\frac{\partial^2 A}{\partial \tilde{\xi}_{1}^{2} }-v_{g} q \left|A\right|^2 A=-i v_{g} \Lambda A,
\label{eq:eq29new}
\end{equation}
which is directly connected to Eq.~(\ref{eq:eq21new}) by a change of the coordinate system.

\begin{figure}
 \centering
 \includegraphics[width=8.5cm]{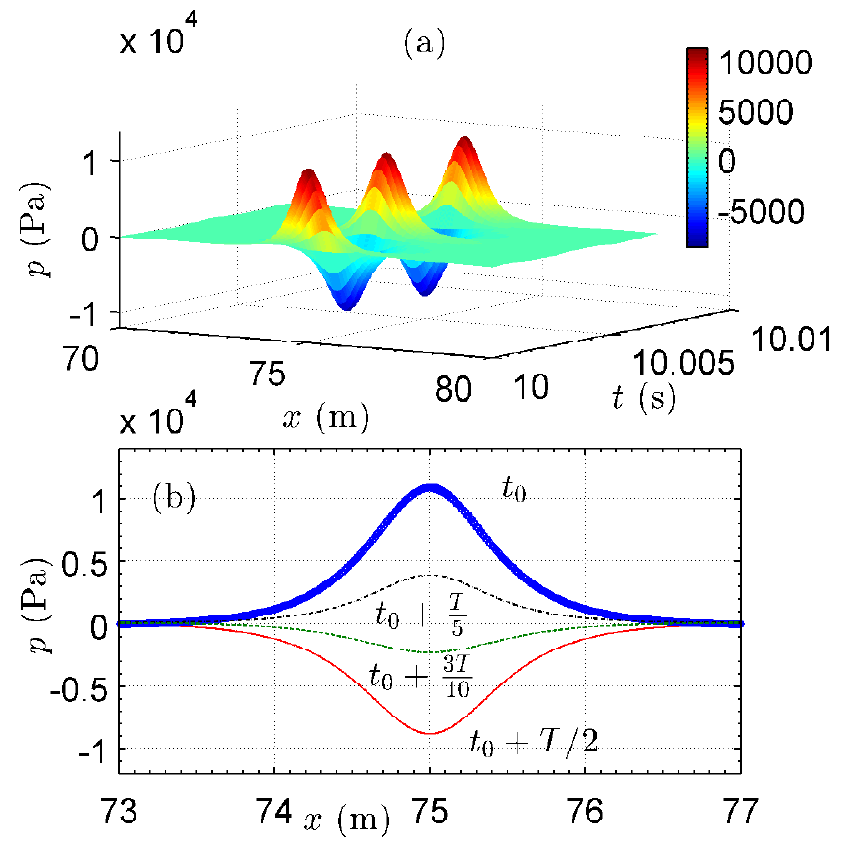} 
  \caption{ (Color online) (a) 3D plot depicting the evolution of a gap soliton of the form of Eq. (\ref{eq:eq36new}), obtained by numerically integrating the lossless version of Eq. (\ref{eq:eq6new}) ($R_{\omega}=0$) with $\epsilon=0.04$ ($2\epsilon\eta P_0=12158$ Pa), in a lattice with a length of $150$ m. (b) Numerical spatial profiles of gap soliton measured from $t_{0}$ (at which gap soliton has a maximal amplitude) to $t_{0}+T/2$ (at which gap soliton has a minimal amplitude).} 
\label{fig:fig5}
\end{figure} 

\begin{figure}[t]
 \includegraphics[width=8.5cm]{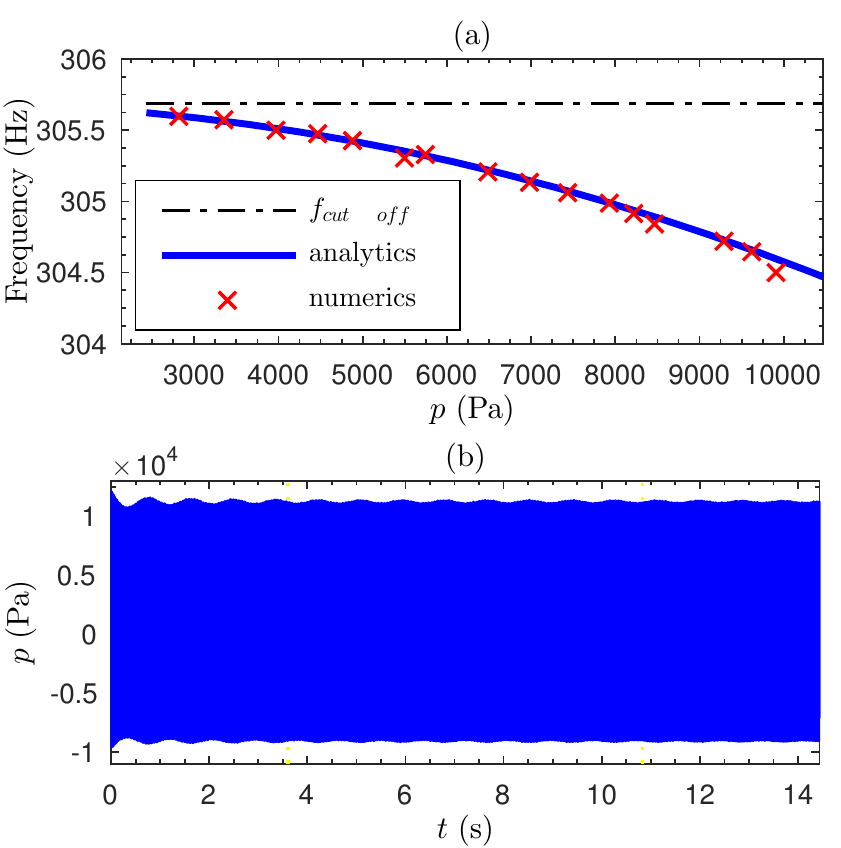} 
 \caption{(Color online) (a) Amplitude dependence of the frequency of the gap soliton. Blue line stands for the analytical results, Eq. (\ref{eq:eq35new}). Red circles stand for the numerical results, where the numerical values of the amplitudes and the frequencies are getting from the main peak of the spectrum of the different gap solitons obtained by numerically integrating the lossless version of Eq. (\ref{eq:eq6new}) ($R_{\omega}=0$) with different initial amplitudes. Black dashed line stands for the cut-off frequency of the system. (b) Time evolution of the middle point of the gap soliton in Fig. \ref{fig:fig5}.}
 \label{fig:fig6}
 \end{figure} 

 \begin{figure*}
 \centering
 \includegraphics[width=16cm]{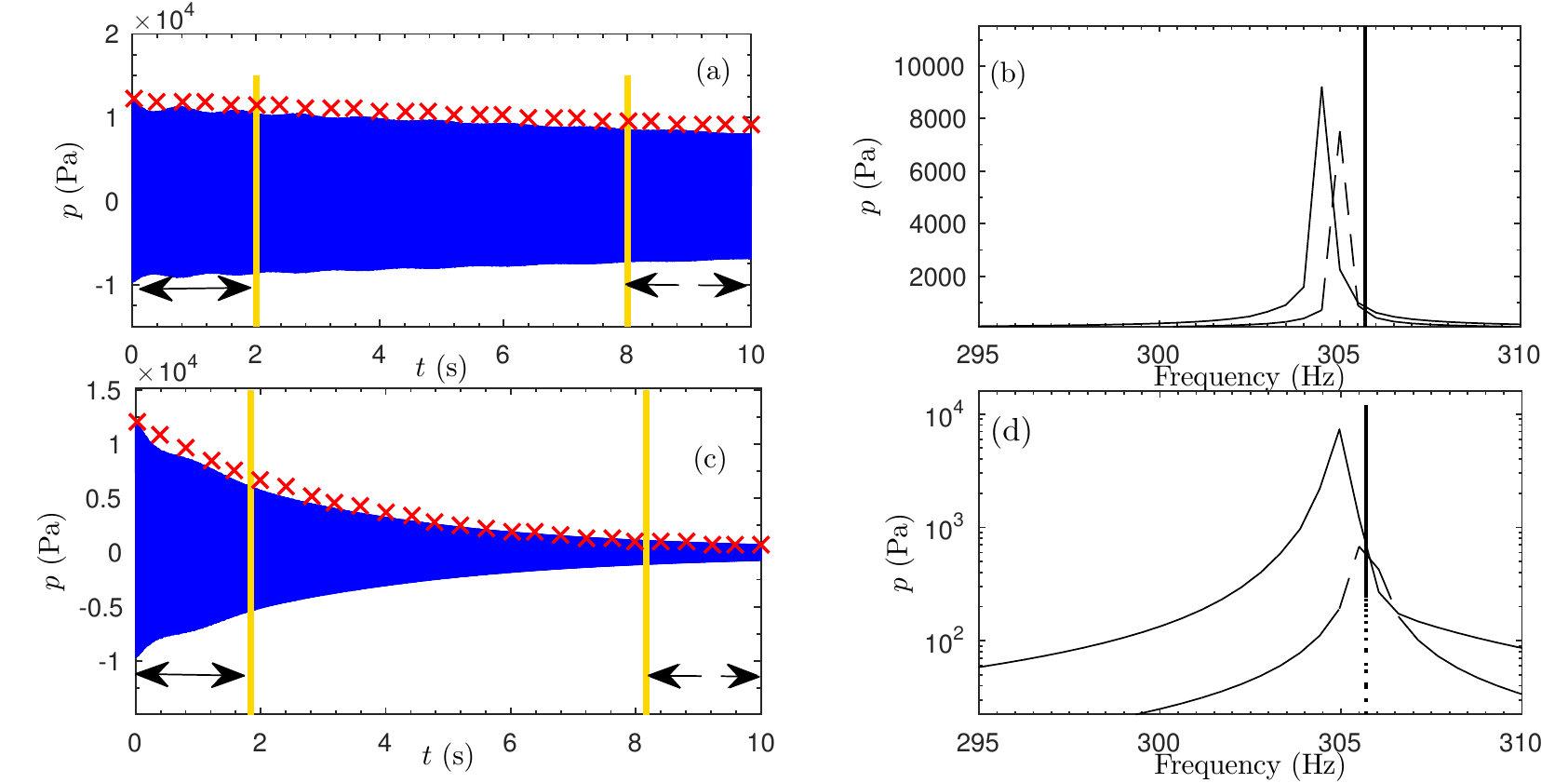} 
  \caption{(Color online) Numerical study 
  %analysis 
  of the effect of viscothermal losses on gap solitons. The Eq. (\ref{eq:eq36new}) with $\epsilon=0.04$ ($2\epsilon\eta P_0=12158$ Pa) is the initial condition for the gap soliton. (a) and (c) represents the time evolution of the middle point of the gap soliton propagating in a weakly lossy medium where $R_{\omega}=6.18$ Ohm and $R_{\omega}=61.8$ Ohm respectively (Blue lines for numerical results and red crosses stand for the analytical time evolution of the maximum pressure for the lossy gap solitons).  (b) and (d) show the spectra of the gap soliton propagating in a lossy medium. Continuous line is the fast Fourier transform (FFT) of the first part of the signal in Fig. \ref{fig:fig9}(a) and Fig. \ref{fig:fig9}(c) and dashed line is the FFT of the last part of the signal in Fig. \ref{fig:fig9}(a) and Fig. \ref{fig:fig9}(c) respectively.} 
  	\label{fig:fig9}
\end{figure*} 

\subsubsection{Gap solitons in the absence of losses}
%Lossless gap solitons} 
In the absence of losses ($\Lambda=0$), the analytical soliton solution for the envelope is of the form,
\begin{equation}
A=\eta \textrm{sech}\left[
\epsilon\eta \sqrt{\left| \frac{q}{k"}\right| }\frac{1}{v_g}(\chi-v_g\tau)\right] \exp(-i \epsilon^2 \eta^2 \frac{q v_{g}}{2} \tau),
\label{eq:eq32new}
\end{equation}
where, as before, $\eta$ 
%is a  parameter setting 
sets the amplitude of the soliton. 

Considering the case with $\omega=m$ and $k=0$, gap soliton solutions of Eq.~(\ref{eq:eq10new}) can then be written in terms of coordinates $\chi$ and $\tau$ as
\begin{equation}
P(\chi, \tau) \approx 2 \eta \textrm{sech}\left(\epsilon \eta \sqrt{\frac{14}{3} }m \beta_0 \chi   \right) %\times 
\cos\left(  \Omega_m \tau \right),
\label{eq:eq33new}
\end{equation}
\noindent
where
\begin{equation}
\Omega_m=m-\frac{7}{3} \epsilon^2 \eta^2 m \beta_{0}^{2}.
\label{eq:eq35new}
\end{equation}
%From the above expressions, it is clear that the gap soliton does not move 
%and it oscillates at a frequency $\Omega_m$, which depends on the amplitude.

In terms of the original space and time coordinates, the approximate envelope gap soliton solution for the pressure $p$ centred at position $x_0$ is the following,
%\begin{widetext}
\begin{eqnarray}
p(x,t)&=&2 \epsilon \eta P_0 \textrm{sech}
\left[ \epsilon \eta \sqrt{\frac{14}{3} }m \beta_0 \omega_B \frac{\sqrt{1+\alpha}}{c_0 }(x-x_0) \right] \nonumber \\
&\times& \cos\left(\Omega_m\omega_B t \right).
\label{eq:eq36new}
\end{eqnarray}
%\end{widetext}
The gap soliton, which is a solution that does not move, is characterized 
by an amplitude $2 \epsilon \eta P_0$. Its width also depends on amplitude 
and it oscillates in time with a period $T=2\pi/\Omega_m\omega_B$.

To study these solutions, we numerically integrate the nonlinear lossless 
lattice model, Eq.~(\ref{eq:eq6new}) with $R_{\omega}=0$, using an initial condition  
given by Eq. (\ref{eq:eq36new}) for $t=0$ and 
%for the gap soliton, localized at 
$x_0=75$ m. An example of a gap soliton with $\epsilon=0.04$ ($2\epsilon\eta P_0=12158$ Pa) is shown in Fig.~\ref{fig:fig5}. %, Fig. \ref{fig:fig6}(a) and Fig. \ref{fig:fig6}(b). 
Figure \ref{fig:fig5}(a) shows the spatio-temporal
 evolution of the gap soliton during a time interval of three periods. Figure \ref{fig:fig5}(b) depicts the numerical spatial profiles of the gap soliton measured from $t_{0}$ (at which gap soliton has a maximal amplitude) to $t_{0}+T/2$. Note that, the absolute value of the maximal amplitude is bigger than that of the minimal amplitude of the gap soliton. 
This asymmetry is caused by the term $\propto |A|^2$ in 
%amplitude shift is due to the DC term in the nonlinear 
Eq.~(\ref{eq:eq20new}). 

%Gap solitons oscillate with a frequency that depends on the amplitude, Eq. (\ref{eq:eq35new}). 
We have calculated both numerically and analytically the frequency of the gap soliton 
for different amplitudes, as shown in Fig.~\ref{fig:fig6}(a). As expected by Eq.~(\ref{eq:eq35new}), 
the frequency of the gap soliton lies in the band gap (blue continuous line);  
red crosses 
%stand for 
depict the numerical results. Each point, represents the frequency of 
the main peak of the spectrum after numerical integration of the lossless version 
of Eq.~(\ref{eq:eq6new}) ($R_{\omega}=0$). It is clearly observed that 
the analytical results 
%(blue line) 
are in a good agreement with the numerical ones. 
%(red crosses) are in a good agreement with the analytical results (blue line). 

The long time evolution of the center of the gap soliton solution is shown in 
Fig.~\ref{fig:fig6}(b). First we note that the amplitude exhibits a long-lived oscillation. 
This can be associated, as in the previous case of the bright solitons, to the birth of 
an internal mode\cite{ref-internalmode1,ref-internalmode2}. 
These beatings are diminished with time as the initial approximate solution 
radiates and approaches the numerically exact gap soliton solution of 
the lattice nonlinear equation.

\subsubsection{Gap solitons in the presence of losses}
%Lossy gap solitons}

We next study numerically the effect of viscothermal losses on the gap soliton.
%As in the previous case for bright solitons, $R_{\omega}(\omega_{g})$ represents the constant losses of the system evaluated at the carrier frequency $\omega_g$ of the gap soliton. 
We numerically integrate Eq.~(\ref{eq:eq6new}) considering  the weak and strong lossy cases, 
as for the bright soliton. The initial condition is of the form of Eq.~(\ref{eq:eq36new}) 
with $t=0$ and 
%localized at 
$x_0=75$ m. We use an amplitude of $\epsilon=0.04$ ($2\epsilon\eta P_0=12158$ Pa) 
and carrier frequency $f=304$ Hz. Figures~\ref{fig:fig9}(a) and \ref{fig:fig9}(b) correspond to the temporal evolution and evolution of the frequency spectrum 
%frequency analysis 
of the amplitude of the gap soliton at $x_0$ in a weakly lossy medium, respectively. 
We observe that the amplitude of the gap soliton decreases slowly with time. As a result, the frequency increases, moving towards the cut-off frequency, see Fig.~\ref{fig:fig9}(b). This is predicted from Eq. (\ref{eq:eq35new}) and illustrated in Fig. \ref{fig:fig6}(a). 

Analogously, we can see in Figs.~\ref{fig:fig9}(c) and \ref{fig:fig9}(d) the 
temporal evolution and frequency spectrum 
%analysis 
of the amplitude of the gap soliton at $x_0$ in a strong lossy medium respectively. In this case, we observe that the amplitude of the gap soliton decays faster than in the weakly lossy medium, 
--see Fig. \ref{fig:fig9}(c)-- and finally its frequency approaches to the cut off frequency. 

Analytical solutions of the lossy problem can also be obtained for the gap solitons. In particular, 
following soliton perturbation theory as before, 
%to obtain 
the evolution of the amplitude of the gap soliton $\eta$ is found to be:
%given by
\begin{equation}
\eta(T_2)=\eta(0)\exp(-2 v_g \Lambda T_2).
\end{equation}
In terms of the original time coordinate, the amplitude of the gap soliton decreases 
exponentially as
\begin{equation}
\eta(t)=\eta(0)\exp(-2 v_g \Lambda \epsilon^2 \omega_B t).
\end{equation}
The analytical results are shown in Figs. \ref{fig:fig9}(a) and \ref{fig:fig9}(c), 
and are found in a good agreement with the numerical results.

\section{Conclusions}\label{sec5}

In conclusion, we have theoretically and numerically studied envelope solitonic structures,   
namely bright and gap solitons, in a 1D acoustic metamaterial composed of an air-filled tube with a periodic array of clamped elastic plates. Based on the electro-acoustic analogy, we 
%proposed 
utilized the transmission line (TL) approach to derive a lossy nonlinear lattice model. 
Considering the continuum limit of the latter, we derived a nonlinear dispersive and dissipative wave equation. In the linear limit, the dispersion relation was found to be in good agreement with the one obtained by the transfer matrix method. 
%There is almost 
No essential difference between the lossy dispersion relation and the lossless one was found, 
because losses are sufficiently small, i.e., the lossy term can be treated as a small perturbation. 

We have thus used a multiple scale perturbative approach to derive an effective NLS model, and
%study 
analytically predict the existence of both bright and gap solitons. The dynamics of these structures  
were studied in the absence and in the presence of viscothermal losses. 
Analytical and numerical results were found to be in very good agreement. 
It is thus concluded that 1D acoustic membrane-type metamaterial can support 
envelope solitary waves even in the presence of realistic viscothermal losses. 
Our results 
%This 
pave the way for the study of nonlinear coherent structures in higher-dimensional settings, 
as well as in double negative metamaterials.

%%%%%%%%%%%%%%%%%%%%%%%%%%%%%%%%%%%%%%%%%%
\vspace{6pt}  %%MDPI internal note: new layout%%
%% optional
%\supplementary{\textbf{Supplementary Materials:} The following are available online at www.mdpi.com/link, Figure S1: title, Table S1: title, Video S1: title.}  %%MDPI internal note: new layout%%

%%%%%%%%%%%%%%%%%%%%%%%%%%%%%%%%%%%%%%%%%%

\begin{acknowledgements} 
Dimitrios J. Frantzeskakis (D.J.F.) acknowledges warm hospitality at Laboratoire d'Acoustique de l'Universit\'e du Maine (LAUM), Le Mans, where most of his work was carried out.
\end{acknowledgements}
%%%%%%%%%%%%%%%%%%%%%%%%%%%%%%%%%%%%%%%%%%

%\authorcontributions{.}%%%Please provide full names.

%%%%%%%%%%%%%%%%%%%%%%%%%%%%%%%%%%%%%%%%%%

\appendix 

\section{Electro-Acoustic Analogue Modeling}\vspace{6pt}

Here, we derive the evolution equation (considering lossy effect of the waveguide) for the pressure in the $n$-th cell of the lattice, as follows. 

First, we note that the advantage of the considered unit-cell circuit is that the inductances $L_{\omega}$ and $L_{m}$ are in a series connection and, thus, can be substituted by the global inductance $L=L_{\omega}+L_{m}$ (see Fig. \ref{fig:fig1}(c)).

Applying Kirchoff's voltage law for two successive cells yields

\begin{equation}
	p_{n-1}-p_{n}=  L \frac{d}{dt}u_{n}+V_{n}+R_{\omega}u_{n},
	\label{eq:eqA1}
\end{equation}

\begin{equation}
	p_{n}-p_{n+1}= L \frac{d}{dt}u_{n+1}+V_{n+1}+R_{\omega}u_{n+1},
	\label{eq:eqA2}
\end{equation}

\noindent
where $V_n$ is the voltage produced by the capacitance of the elastic plates $C_{m}$. Subtracting the two equations above, we obtain the differential-difference equation (DDE)

\begin{equation}
	\hat{\delta}^{2} p_{n}= L \frac{d}{dt}\left(u_{n}-u_{n+1}\right)+R_{\omega}\left(u_{n}-u_{n+1}\right)+\left(V_{n}-V_{n+1}\right)
	\label{eq:eqA3},
\end{equation}

\noindent
where $\hat{\delta}^{2} p_{n}\equiv p_{n+1}-2p_{n}+p_{n-1}$. Then, Kirchhoff's current law yields

\begin{equation}
	u_{n}-u_{n+1}=C_{\omega}\frac{d}{dt}\left(p_{n}\right)
	\label{eq:eqA4},
\end{equation}

\noindent
with 

\begin{equation}
	u_{n}=C_{m}\frac{d}{dt}\left(V_{n}\right) \quad \mbox{and} \quad u_{n+1}=C_{m}\frac{d}{dt}\left(V_{n+1}\right)
	\label{eq:eqA5}.
\end{equation}

Subtracting Eq. (\ref{eq:eqA5}) and employing Eq. (\ref{eq:eqA4}), we obtain

\begin{equation}
	u_{n}-u_{n+1}=C_{m}\frac{d}{dt}\left(V_{n}-V_{n+1}\right)=C_{\omega}\frac{d}{dt}\left(p_{n}\right).
	\label{eq:eqA6}
\end{equation}

\noindent
Then, recalling that the capacitance $C_{\omega}$ depends on the pressure (cf. Eq. (\ref{eq:eq4new})), we express $V_{n}-V_{n+1}$ as

\begin{equation}
	V_{n}-V_{n+1}=\frac{C_{\omega}}{C_{m}}p_{n}=\frac{C_{\omega 0}}{C_{m}}p_{n}-\frac{C_{\omega}^{'}}{C_{m}}p_{n}^{2}.
	\label{eq:eqA7}
\end{equation}

Next, substituting Eq. (\ref{eq:eqA6}) and Eq. (\ref{eq:eqA7}) into Eq. (\ref{eq:eqA3}), 
we obtain the following evolution equation for the pressure

\begin{equation}
	\hat{\delta}^{2} p_{n}=L \frac{d}{dt}\left(C_{\omega}\frac{d}{dt}\left(p_{n}\right)\right)+R_{\omega}\left(C_{\omega}\frac{d}{dt}\left(p_{n}\right)\right)+\frac{C_{\omega}}{C_{m}}p_{n}.
	\label{eq:eqA8}
\end{equation}

To this end, employing Eq. (\ref{eq:eq4new}), we can rewrite the above equation, then we get the evolution equation (considering lossy effect of the waveguide) for the pressure in the $n$-th cell of the lattice, Eq. (\ref{eq:eq6new}).

%\appendix %%MDPI internal note: new layout%%
%\setcounter{equation}{0}
%\renewcommand\theequation{B\arabic{equation}}
%
%\setcounter{figure}{0}
%\renewcommand\thefigure{B \arabic{figure}} 

\section{Hierarchy of equations in multiple scale perturbation method}\vspace{6pt}

There we present the hierarchy of equations at various orders in $\epsilon$,

\begin{equation}
   O(\epsilon^0) : \widehat{L}_0 p_0=0,
   	\label{eq:eqB1}
   \end{equation}

 \begin{equation}
   O(\epsilon^1) : \widehat{L}_0 p_1+\widehat{L}_1 p_0=\widehat{N}_0 \left[ p_{0}^{2} \right],
      	\label{eq:eqB2}
   \end{equation}

\begin{equation}
   O(\epsilon^2) : \widehat{L}_0 p_2+\widehat{L}_1 p_1+\widehat{L}_2 p_0=\widehat{N}_0 \left[ 2p_{0}p_{1} \right]+\widehat{N}_1 \left[ p_{0}^{2} \right],
      	\label{eq:eqB3}
   \end{equation}

\noindent
where linear operators $\widehat{L}_0$, $\widehat{L}_1$ and $\widehat{L}_2$, as well as the nonlinear operators $\widehat{N}_0$, $\widehat{N}_1$ are given by

\begin{equation}
\widehat{L}_0=-\frac{\partial^2}{\partial \chi_{0}^{2}}+\frac{\partial^2}{\partial \tau_{0}^{2}}-\zeta\frac{\partial^4}{\partial \chi_0^4} +m^2,
\label{eq:eqB4}
\end{equation}

\begin{equation}
\widehat{L}_1=-2\frac{\partial^2}{\partial \chi_{0} \chi_{1}}+2\frac{\partial^2}{\partial \tau_{0} \tau_{1}}-4\zeta\frac{\partial^4}{\partial \chi_0^3 \partial \chi_1} ,
\label{eq:eqB5}
\end{equation}

\begin{equation}
\begin{aligned}
\widehat{L}_2= & -\frac{\partial^2}{\partial \chi_{1}^{2}}-2\frac{\partial^2}{\partial \chi_{0} \chi_{2}}+\frac{\partial^2}{\partial \tau_{1}^{2}}+2\frac{\partial^2}{\partial \tau_{0} \tau_{2}}\\
&-\zeta\left(6\frac{\partial^4}{\partial \chi_0^2 \partial \chi_1^2}+4\frac{\partial^4}{\partial \chi_0^3 \partial \chi_2}\right)+\Gamma \frac{\partial }{\partial \tau_0},
\end{aligned}
\label{eq:eqB6}
\end{equation}
 
\begin{equation}
\widehat{N}_0=\beta_0 \frac{\partial^2}{\partial \tau_{0}^{2}}+2 \beta_0 m^2,
\label{eq:eqB7}
\end{equation}
 
\begin{equation}
\widehat{N}_1=2 \beta_0 \frac{\partial^2}{\partial \tau_{0} \tau_{1}}.
\label{eq:eqB8}
\end{equation}

%\appendix %%MDPI internal note: new layout%%
%\setcounter{equation}{0}
%\renewcommand\theequation{C\arabic{equation}}
%
%
%\setcounter{figure}{0}
%\renewcommand\thefigure{C \arabic{figure}} 

\section{Nonlinear length and dispersion length}\vspace{6pt}

Here is the calculation for nonlinear length and dispersion length.

We can rewrite Eq. (\ref{eq:eq21new}) in its dimensional form as

\begin{equation}
i\frac{\partial \phi}{\partial x}-\frac{1}{2} k_{ph}^{''}\frac{\partial^2 \phi}{\partial T^{2} }-q_{ph} \left|\phi\right|^2 \phi=0,
\label{eq:eqC1}
\end{equation}

\noindent
where 

\begin{equation}
k_{ph}^{''}=\frac{k''}{\omega_B c},
\quad
q_{ph}=q(  \omega, k) \frac{\omega_{B}}{c}\frac{1}{P_0^2}, 
\label{eq:eqC2}
\end{equation}

\noindent
and $\phi /P_0=  \epsilon  A $, $T=t-x/v_g$, $v_g=\partial  \omega_{ph}/\partial k_{ph}$.

In order to get the dispersion length and the nonlinearity length, we introduce $t_0$ and $A_0$ as the characteristic width of the initial condition, and the maximum pressure of the initial condition respectivelly. Then we use the new time variable $\tilde{T}=T/t_0$ and substitute $\phi=A_0 \Phi $ to obtain

\begin{equation}
i\frac{\partial \Phi}{\partial x}-\frac{1}{2L_D} \frac{\partial^2 \Phi}{\partial \tilde{T}^2 }-\frac{1}{L_{NL}} \left|\Phi\right|^2 \Phi=0,
\label{eq:eqC3}
\end{equation}

\noindent
where the characteristic lengths are defined as,

\begin{equation}
L_D=\frac{t_0^2}{ \left| k_{ph}^{''} \right| }, \mbox{ and}
L_{NL}=\frac{1}{\left| q_{ph} \right|   A_0^2    }.
\label{eq:eqC4}
\end{equation}

According to Eq. (\ref{eq:eq24new}), here we define,

\begin{equation}
t_0=\left(\epsilon \eta \sqrt{\left| \frac{q}{k^{''}}\right| } \omega_B \right)^{-1}, 
\quad  \mbox{ and }
A_0=\epsilon \eta P_0.
\label{eq:eqC5}
\end{equation}

Thus $L_{NL}/L_D \sim 1$.

%\bibliography{references}


\end{document}